\documentclass[conference]{IEEEtran}

\usepackage{amsmath,amssymb}
\usepackage{booktabs}
\usepackage{multirow}
\usepackage{graphicx}
\usepackage{xcolor}
\usepackage{cite}
\usepackage{url}
\usepackage{algorithm}
\usepackage{algpseudocode}
\usepackage{tikz}
\usepackage{pgfplots}
\pgfplotsset{compat=1.18}
\usetikzlibrary{arrows.meta,positioning,fit,shapes.geometric,calc}

\definecolor{goodgreen}{RGB}{223,242,225}
\definecolor{warnyellow}{RGB}{255,244,204}
\definecolor{badred}{RGB}{250,220,220}
\definecolor{coolblue}{RGB}{222,235,250}
\definecolor{darkblue}{RGB}{34,83,133}

\title{A Reference-Free Score for Detecting Silent\\
Reasoning Failures in Large Language Models}

\newcommand{\authorcell}[2]{%
\parbox[t][6.7em][t]{0.29\textwidth}{\centering
{\fontsize{10.5}{12}\selectfont #1}\par\vspace{0.3em}
{\fontsize{8.5}{10.5}\selectfont
\textit{Allenhouse Institute of Technology}\\[0.25em]
\textit{Kanpur, India}\\[0.25em]
#2}}}

\author{%
\makebox[\textwidth][c]{%
\setlength{\tabcolsep}{0.8em}%
\begin{tabular}{ccc}
\authorcell{Vivek Shukla}{vivekshukla0552@gmail.com}
&
\authorcell{Varun Shukla}{varun.shuklaa@gmail.com}
&
\authorcell{Atul}{atulverma15704@gmail.com}
\\[1.2em]
\authorcell{Divya Mishra}{divyamishra03125@gmail.com}
&
\authorcell{Mehul Kumar Das}{dasmehulkumar08@gmail.com}
&

\end{tabular}}}

\begin{document}
\maketitle

\begin{abstract}
Mathematical chain-of-thought (CoT) evaluation is commonly reduced to whether the final answer matches a reference. This conflates producing a correct conclusion with producing a valid derivation: an invalid chain can accidentally reach the right answer, while a valid calculation can be followed by a transcription error. We call this mismatch the \emph{reasoning--answer consistency gap}. This framework paper introduces the \emph{Reasoning--Answer Faithfulness Score} (RAFS), a reference-free, instance-level diagnostic of whether an emitted mathematical trace is locally credible, supports its answer, and is stable under resampling and targeted counterfactual interventions. RAFS combines step validity, reasoning-to-answer entailment and counterfactual sensitivity, answer consensus, and conditional reasoning stability. It evaluates transcript-level agreement, not a model's private computation and not factual correctness outside the tested mathematical setting. We retain a preregistered, results-blind confirmatory study on GSM8K and MATH, with hypotheses, admissibility rules, calibration, and tests fixed before confirmatory outcomes are inspected. A separate feasibility pilot is specified to verify end-to-end execution and estimate intervention coverage before that freeze; numerical pilot claims are reported only when trace-level artifacts are available. We formalize four reasoning--answer outcomes, justify the non-compensatory aggregator, instantiate semantic trace distance, quantify compute and abstention tradeoffs, and define verifier-independence and power analyses. RAFS is intended to complement mathematical answer accuracy with an auditable warning signal for silent reasoning failures and answer-extraction errors
\end{abstract}

\begin{IEEEkeywords}
chain-of-thought, reasoning evaluation, faithfulness, uncertainty, self-consistency, explainable artificial intelligence, large language models
\end{IEEEkeywords}

\section{Introduction}
Chain-of-thought prompting can improve performance on multi-step tasks by eliciting intermediate natural-language calculations or deductions before a final answer \cite{wei2022cot}. The same traces are also increasingly presented as explanations: a user is invited to trust not only the answer, but the apparent path by which it was obtained. These two roles are not equivalent. A trace can be useful for generation yet fail to be a valid justification, and a fluent justification can be produced after an answer has effectively been selected \cite{turpin2023unfaithful}. Treating final-answer accuracy as a proxy for reasoning quality therefore creates a consequential blind spot.

Consider a model that makes two compensating arithmetic errors and nevertheless outputs the reference answer. Exact-match accuracy deems a response as correct even though the derivation would be incorrect with a slight perturbation. This falls under the category of reasoning false positive. The answer is correct, but the reasoning is incorrect. We refer to such instances as silent reasoning failures. Outcome-only evaluations disguise the reasoning. On the other hand, a model may perform an accurate derivation, but copy the incorrect answer from the multiple-choice options or perform an incorrect rounding in the final step. This would be classed as a reasoning false negative. Even though most of the reasoning is correct, the answer is still incorrect. Combining the two axes into one bit obscures the models capability, and gives poor training feedback, while possibly making an incorrect output appear correct, which is a dangerous position to be in.

This paper restricts its claims to mathematical reasoning, where arithmetic and algebraic steps are comparatively auditable. Even in this restricted setting, the hidden defect may reverse an answer under a small change to the problem, while rejecting a sound calculation because of formatting misidentifies the component requiring repair. Existing accuracy benchmarks cannot distinguish these cases; answer self-consistency aggregates outcomes without necessarily validating their chains, and step verifiers may judge each step without measuring whether the final answer follows from the trace.

The confirmatory GSM8K/MATH study remains preregistered and results-blind: its hypotheses, model families, annotation protocol, intervention rules, calibration choices, baselines, and tests are fixed before confirmatory outcomes are examined. A small, disjoint feasibility pilot may be used only to establish that the pipeline runs, measure intervention coverage and cost, and debug implementation. It may not be used to select confirmatory hypotheses or report confirmatory performance. boundary retains falsifiability and lets operational gaps be seen before an expensive annotation study.The Reasoning-Answer Faithfulness Score (RAF score) is an automatic measure of [0, 100] that reflects the extent to which generated reasoning trace R and final answer A are in agreement, and is calculated without gold answer requirement for system deployment. RAF score considers local process credibility, global answer support, resampled answer consensus, and reasoning stability. Strategic intervention can determine if changing a certain pivotal step leads to changes in the answer to a directionally expected form, therefore, the score on polished, causally inert explanations is reduced. RAF score should not be seen as a providence oracle since high agreement, in multiple sampled reasoning traces, can occur with the same
misconception. It is more a fair, calibrated assessment of the observable answer and explanation.The innovation is threefold:
\begin{itemize}
    \item Reasoning correctness and answer correctness are construed as distinct dimensions which allows the disagreement of the two, rather than answer error, to take precedence in the evaluation.
    \item A RAF score is an integrated, reference-free metric of support that incorporates step-wise validation, counterfactual sensitivity,and a stretch of reasoning uncertainty.
    \item A score, type of failure, the earliest suspect step,and an abstention decision are given with a deployment focused detector, allowing verification or regeneration requests, versus accepted answer in a binary sense.
\end{itemize}

\section{Related Work}
\subsection{Outcome-Based and Self-Consistent Reasoning}
CoT prompting and self-consistency decoding showed that sampling and aggregating multiple reasoning pathways can improve final-answer accuracy \cite{wang2023selfconsistency}. Thus, GSM8K and MATH became popular benchmarks for assessing mathematical reasoning. However, exact match only compares the answer $A$ with a reference, and majority voting only assesses agreement among answers. Neither establishes that an individual reasoning trace $R$ is valid or that $A$ follows from $R$. Dataset contamination and template memorization further obscure the interpretation of high benchmark scores \cite{zhang2024gsm1k}.

\subsection{Faithfulness and Reasoning-Trace Evaluation}
Prior work shows that CoT explanations can omit influential prompt features or rationalize biased answers. Intervention-based tests such as truncating or corrupting a chain probe whether later predictions depend on earlier reasoning \cite{lanham2023faithfulness}. Parcalabescu and Frank argue that many purported faithfulness tests more precisely measure output-level self-consistency, not correspondence to inaccessible internal computation \cite{parcalabescu2024selfconsistency}. We adopt this distinction: RAFS measures \emph{observable reasoning--answer faithfulness} and uses interventions as evidence, not proof, of causal mediation.

Reference-free metrics score semantic alignment, logicality, informativeness, and fluency, while direct evaluation frameworks reconstruct or judge reasoning paths. These approaches improve process visibility, but they do not jointly encode the four-quadrant outcome taxonomy and the stability of the $R\!\rightarrow\!A$ relation across samples.

\subsection{2024--2025 Process Supervision}
Recent work increasingly evaluates steps rather than outcomes. Causal objectives and inferential bridging analyze the relevance among context, CoT, and answer. ProcessBench provides expert labels for locating the earliest error in mathematical solutions and reports limited generalization of several process reward models (PRMs) to harder problems \cite{zheng2024processbench}. ThinkPRM instead performs generative, step-wise verification with substantially fewer process labels than conventional discriminative PRMs. Studies of contemporary reasoning models further suggest that explicit CoTs may fail to disclose prompt influences and may be unreliable monitoring targets.

RAFS complements these directions. Unlike a PRM, it is not a new reward model architecture; unlike an outcome verifier, it does not reduce a trace to answer likelihood; unlike a similarity-only metric, it tests whether pivotal reasoning changes propagate to $A$. Its contribution is a model-agnostic aggregation protocol and diagnostic taxonomy that can wrap a critic model, symbolic checker, NLI model, or specialized PRM.

\begin{table*}[t]
\caption{Component-level comparison. A checkmark denotes an explicit target of the method; ``partial'' denotes related evidence that does not instantiate the RAFS component.}
\label{tab:related-components}
\centering
\footnotesize
\begin{tabular}{@{}lccccp{6.0cm}@{}}
\toprule
Method & $P_R$ & $S_{RA}$ & $C_A$ & $D_R$ & Primary object measured \\
\midrule
ProcessBench & \checkmark & -- & -- & -- & Expert benchmark for step-error detection and earliest-error localization. \\
ThinkPRM & \checkmark & -- & -- & -- & Generative step-wise process verification; it can supply $q_t$ but does not test answer dependence. \\
ROSCOE & partial & partial & -- & partial & Reference-free semantic alignment, logicality, informativeness, and fluency; no controlled $R\!\rightarrow\!A$ intervention. \\
FRODO & -- & \checkmark & -- & -- & Causal/counterfactual rationale influence as a training and evaluation objective. \\
RAFS (ours) & \checkmark & \checkmark & \checkmark & \checkmark & Instance-level aggregation plus four-quadrant diagnosis, coverage-aware abstention, and suspect-step output. \\
\bottomrule
\end{tabular}
\end{table*}

\section{Problem Formulation}
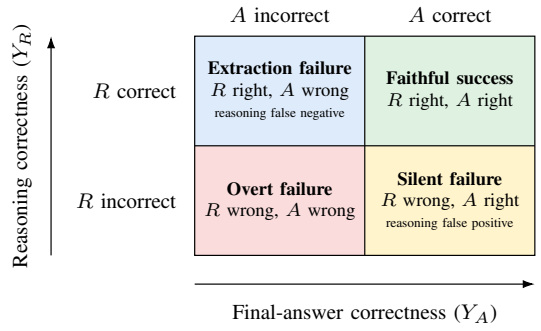
\begin{figure}[t]
\centering
\begin{tikzpicture}[font=\scriptsize, x=1cm,y=1cm]
  \fill[coolblue]   (0,1.45) rectangle (2.25,2.9);
  \fill[goodgreen]  (2.25,1.45) rectangle (4.5,2.9);
  \fill[badred]     (0,0) rectangle (2.25,1.45);
  \fill[warnyellow] (2.25,0) rectangle (4.5,1.45);

  \draw[line width=.45pt] (0,0) rectangle (4.5,2.9);
  \draw[line width=.45pt] (2.25,0)--(2.25,2.9);
  \draw[line width=.45pt] (0,1.45)--(4.5,1.45);

  \node[font=\footnotesize] at (1.125,3.18) {$A$ incorrect};
  \node[font=\footnotesize] at (3.375,3.18) {$A$ correct};
  \node[font=\footnotesize,anchor=east] at (-.12,2.18) {$R$ correct};
  \node[font=\footnotesize,anchor=east] at (-.12,.72) {$R$ incorrect};

  \node[align=center] at (1.125,2.18)
    {\textbf{Extraction failure}\\$R$ right, $A$ wrong\\{\tiny reasoning false negative}};
  \node[align=center] at (3.375,2.18)
    {\textbf{Faithful success}\\$R$ right, $A$ right};
  \node[align=center] at (1.125,.72)
    {\textbf{Overt failure}\\$R$ wrong, $A$ wrong};
  \node[align=center] at (3.375,.72)
    {\textbf{Silent failure}\\$R$ wrong, $A$ right\\{\tiny reasoning false positive}};

  \draw[-{Latex[length=1.6mm,width=1.1mm]},line width=.45pt]
    (0,-.38)--(4.5,-.38);
  \node[font=\footnotesize,below=1.5mm] at (2.25,-.38)
    {Final-answer correctness ($Y_A$)};
  \draw[-{Latex[length=1.6mm,width=1.1mm]},line width=.45pt]
    (-1.90,0)--(-1.90,2.9);
  \node[font=\footnotesize,rotate=90] at (-2.28,1.45)
    {Reasoning correctness ($Y_R$)};
\end{tikzpicture}
\caption{The reasoning--answer consistency gap. Accuracy observes only the horizontal axis and therefore merges faithful success with silent failure, and extraction failure with overt failure.}
\label{fig:quadrant}
\end{figure}

Let $x$ be a problem, $R=(r_1,\ldots,r_T)$ a natural-language reasoning trace divided into $T$ atomic steps, and $A$ the extracted final answer. When references are available for evaluation, let $A^*$ denote the gold answer and $V^*\in\{0,1\}$ denote expert judgment that the chain is a valid derivation for $x$. Define answer correctness $Y_A=\mathbb{1}[A\equiv A^*]$ and reasoning correctness $Y_R=V^*$. The pair $(Y_R,Y_A)$ induces four outcomes (Fig.~\ref{fig:quadrant}).

This separation is useful for formal mathematical and logical tasks in machine
learning \cite{awasthi2024mathlogic}, including structured algebraic settings
in which intermediate polynomial operations must remain auditable
\cite{shukla2021auth}.

The evaluation objective is not merely to estimate $Y_A$, but to estimate whether $R$ is trustworthy and supports $A$. At deployment, $A^*$ and $V^*$ are unavailable. We therefore construct observable proxy signals from a verifier ensemble, sampled solutions, and controlled interventions. RAFS estimates the consistency event
\begin{equation}
F(x,R,A)=\mathbb{1}[R\text{ is credible}]\,\mathbb{1}[R\models A],
\end{equation}
while separately retaining uncertainty. On labeled test sets, RAFS is evaluated against $Y_R=Y_A$ and, more importantly, against the two off-diagonal error classes.

\section{Reasoning--Answer Faithfulness Score}
\subsection{Step Validity}
An atomic-step verifier $g$ receives $(x,r_{1:t})$ and returns
\begin{equation}
q_t=P_g(v_t=1\mid x,r_{1:t}),
\end{equation}
where $v_t$ indicates that step $t$ follows from the problem and preceding steps. The verifier may combine an LLM critic, deterministic arithmetic checks, contradiction detection, correctness/informativeness evaluation, and verifier-benchmark supervision \cite{prasad2023receval,jacovi2024reveal}. To penalize a single fatal step while remaining differentiable, we use a mixture of geometric mean and minimum:
\begin{equation}
P_R=\rho\exp\left(\frac{1}{T}\sum_{t=1}^{T}\log(q_t+\epsilon)\right)
 +(1-\rho)\min_t q_t .
\label{eq:process}
\end{equation}
The earliest $t$ below a calibrated threshold is returned as the suspect step.

\subsection{Reasoning-to-Answer Support}
Let $e=P_g(R\models A\mid x)$ be the verifier's probability that the final answer is entailed by the complete trace. Entailment alone is vulnerable to post-hoc rationalization. We therefore identify a set $J$ of pivotal steps using leave-one-step-out influence or verifier attribution, and produce candidate counterfactual chains $R^{(j)}$ that alter the conclusion of step $j$. This intervention design is motivated by causal-rationale and inferential-bridging approaches \cite{paul2024frodo,li2024bridging}. The intended perturbation direction (e.g., increasing a computed quantity or reversing a Boolean claim) is recorded before the answer head is queried. Let $A^{(j)}$ be the answer obtained when an answer head reads an admissible $R^{(j)}$, and let $\Delta_j\in[0,1]$ measure whether the answer changes in that preregistered direction.

We distinguish \emph{intervention validity} from whether the edited claim is true: a useful counterfactual may deliberately make the target claim false, but it must isolate that change. A candidate is admitted only if it passes all of the following preregistered rubric items:
\begin{enumerate}
    \item \textbf{Atomicity and locality:} $r_j$ contains one designated proposition $c_j$, and the edit changes only $c_j$. All steps $r_i$ for $i\neq j$, all non-target propositions within $r_j$, and the problem $x$ are held fixed; the original answer span is masked from the edit generator.
    \item \textbf{Bounded form change:} after answer-span masking and token normalization, the normalized token-level Levenshtein distance satisfies
    \begin{equation}
        d_{\mathrm{edit}}(r_j,r_j^{(j)})=
        \frac{\operatorname{Lev}(r_j,r_j^{(j)})}
        {\max(|r_j|,|r_j^{(j)}|)}\leq\delta_{\mathrm{edit}},
        \label{eq:editbound}
    \end{equation}
    where $\delta_{\mathrm{edit}}$ is fixed in the preregistration (default $0.20$). The edit may not insert a new premise, justification, or solution path.
    \item \textbf{Semantic isolation:} a claim differencer must identify exactly one changed proposition, $c_j\rightarrow c_j'$. The replacement preserves entities, units, variable bindings, quantifiers not under intervention, and the claim's semantic type (numeric, relational, Boolean, or categorical); it introduces no collateral contradiction beyond the intended change to $c_j$ and its direct implications.
    \item \textbf{Well-formedness and non-leakage:} the edited step is grammatical, type- and unit-consistent, and interpretable in the original context. It may not mention, paraphrase, or otherwise leak the expected counterfactual answer.
    \item \textbf{Independent validation:} a symbolic checker is used when the claim is executable; otherwise two validators, not used to generate the edit, judge locality and semantic validity. Both must accept. Disagreement, parser failure, or an ambiguous predicted answer direction rejects the candidate.
\end{enumerate}
The claim differencer and validators operate before $A^{(j)}$ is observed. This ordering prevents an intervention from being retained because it happened to produce the desired response. Let $J_{\mathrm{val}}\subseteq J$ be the steps with admitted interventions. Counterfactual sensitivity is
\begin{equation}
I_{RA}=\frac{1}{|J_{\mathrm{val}}|}\sum_{j\in J_{\mathrm{val}}}\Delta_j,
\qquad
S_{RA}=\lambda e+(1-\lambda)I_{RA}.
\label{eq:support}
\end{equation}
A chain that verbally entails $A$ but has no measurable effect on answer generation receives limited support. Rejected interventions are logged with a failure code and are not counted as negative evidence. If fewer than the preregistered minimum $m_{\min}$ interventions are admissible (default $m_{\min}=2$), $I_{RA}$ is marked unavailable and RAFS abstains rather than imputing counterfactual support.

\subsection{Consensus and Conditional Stability}
Sample $K$ independent pairs $\{(R_k,A_k)\}_{k=1}^{K}$ at nonzero temperature. For normalized answer classes $\mathcal{A}$, let $p(a)=K^{-1}\sum_k\mathbb{1}[A_k=a]$. Normalized answer consensus is
\begin{equation}
C_A=1-\frac{H(p)}{\log(\max(2,|\mathcal{A}|))},
\quad H(p)=-\sum_{a\in\mathcal{A}}p(a)\log p(a).
\label{eq:consensus}
\end{equation}
High answer consensus can coexist with diverse or contradictory explanations. Semantic reasoning metrics motivate comparing normalized claims rather than surface wording \cite{golovneva2023roscoe}. We therefore compute conditional reasoning stability among traces supporting the modal answer $\hat a=\arg\max_a p(a)$. Concretely, $z(R)$ is the ordered multiset of atomic claims after canonicalizing numbers, units, and variable names. For claims $c,c'$, a frozen DeBERTa-large model fine-tuned on MNLI \cite{he2021deberta} supplies directional entailment probabilities, and $d_c(c,c')=1-\tfrac12[p(c\Rightarrow c')+p(c'\Rightarrow c)]$. We define $d(z(R_i),z(R_j))$ as the minimum-cost monotone alignment of their claim sequences under $d_c$, with cost one for an unmatched claim, normalized by the longer sequence. Bidirectional entailment distinguishes contradiction or a missing proof obligation from paraphrase; cosine embedding distance is cheaper but can place contradictory claims close together. Executable numeric claims are canonicalized by symbolic equivalence before NLI scoring. With this fixed $d\in[0,1]$,
\begin{equation}
D_R=1-\frac{2}{|\mathcal{K}_{\hat a}|(|\mathcal{K}_{\hat a}|-1)}
\sum_{i<j\in\mathcal{K}_{\hat a}}d(z(R_i),z(R_j)).
\label{eq:stability}
\end{equation}
If fewer than two traces support $\hat a$, we set $D_R=0$ and force abstention.

\subsection{Composite Score and Calibration}
For weights $\boldsymbol{w}=(w_P,w_S,w_C,w_D)$ on the simplex, RAFS is the weighted geometric mean
\begin{equation}
\mathrm{RAFS}=100\exp\left(\sum_{m\in\{P,S,C,D\}}w_m
\log(s_m+\epsilon)\right),
\label{eq:rafs}
\end{equation}
where $(s_P,s_S,s_C,s_D)=(P_R,S_{RA},C_A,D_R)$. The geometric form is intentionally non-compensatory: excellent fluency or consensus cannot fully erase a near-zero validity or support score. We fit $\rho$, $\lambda$, $\boldsymbol{w}$, and decision thresholds on a held-out development split using expert four-quadrant labels. Calibration uses isotonic regression or temperature scaling, never the test labels.

The aggregator is chosen to match an explicit failure semantics: every
positively weighted component represents necessary evidence. Consequently,
$s_m\rightarrow0$ must imply RAFS$\rightarrow0$. An arithmetic mean violates
this boundary condition because three strong components can compensate for a
failed process-validity or reasoning-support signal. Hard minimum gating
satisfies the boundary condition, but it is non-smooth and allows a single
miscalibrated component to dominate the decision. Under continuity,
monotonicity, the zero-boundary condition, and constant relative sensitivity
$\partial\log f/\partial\log s_m=w_m$, integration yields
$\log f=\sum_m w_m\log s_m+c$. The normalization
$f(1,\ldots,1)=1$ then gives Eq.~\eqref{eq:rafs}.

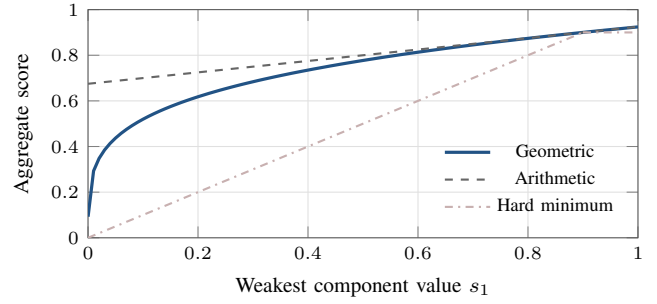
\begin{figure}[t]
\centering
\begin{tikzpicture}
\begin{axis}[
  width=\columnwidth,
  height=4.6cm,
  xmin=0,xmax=1,
  ymin=0,ymax=1,
  xlabel={Weakest component value $s_1$},
  ylabel={Aggregate score},
  tick label style={font=\scriptsize},
  label style={font=\footnotesize},
  legend style={font=\scriptsize,draw=none,fill=none,
    at={(0.98,0.05)},anchor=south east},
  grid=major,
  major grid style={draw=black!12},
  axis line style={black!65},
  samples=101]
  \addplot[darkblue,very thick,domain=0.0001:1]
    {(x*0.9^3)^(1/4)};
  \addlegendentry{Geometric}
  \addplot[black!60,thick,dashed,domain=0:1]
    {(x+2.7)/4};
  \addlegendentry{Arithmetic}
  \addplot[badred!80!black,thick,dash dot,domain=0:0.9] {x};
  \addplot[badred!80!black,thick,dash dot,domain=0.9:1,
    forget plot] {0.9};
  \addlegendentry{Hard minimum}
\end{axis}
\end{tikzpicture}
\caption{Analytical response of candidate aggregators when
$s_2=s_3=s_4=0.9$ and weights are equal. The arithmetic mean remains high
when one necessary signal fails; minimum gating is maximally sensitive; the
geometric mean provides smooth, non-compensatory behavior.}
\label{fig:aggregators}
\end{figure}

Figure~\ref{fig:aggregators} illustrates this distinction by varying one weak
component while fixing the remaining three components at $0.9$ and assigning
equal weights. The arithmetic mean remains approximately $0.68$ even when the
weak component is zero, demonstrating undesirable compensation. Hard minimum
gating follows the weakest component exactly until it reaches $0.9$, making it
maximally sensitive to a single estimate. The geometric mean approaches zero
with the failed component but changes smoothly, which is the intended RAFS
behavior. Arithmetic, minimum-gated, and learned logistic aggregators remain
preregistered ablations rather than being excluded by construction.

Two calibrated thresholds define the deployment states: \emph{accept} when
RAFS $\geq\tau_h$ and uncertainty is low; \emph{review/regenerate} when RAFS
$\leq\tau_l$; and \emph{abstain} otherwise. Without a gold answer, high answer
consensus paired with low $P_R$ or $S_{RA}$ is flagged as a silent-failure
candidate. Low consensus with high $P_R$ and high entailment instead suggests
answer-extraction or decoding instability.

\begin{algorithm}[t]
\caption{Reference-Free RAFS Calculation}
\label{alg:rafs}
\begin{algorithmic}[1]
\Require problem $x$, primary pair $(R,A)$, samples $K$, verifier $g$
\Ensure score $s$, state, suspect step $t^*$
\State segment $R$ into atomic steps $(r_1,\ldots,r_T)$
\For{$t=1$ to $T$}
  \State $q_t\gets g(x,r_{1:t})$; run applicable symbolic checks
\EndFor
\State compute $P_R$ using Eq.~\eqref{eq:process}; $t^*\gets\min\{t:q_t<\tau_v\}$
\State $e\gets g(R\models A\mid x)$; identify pivotal steps $J$
\ForAll{$j\in J$}
  \State generate candidate $R^{(j)}$ and record the predicted response direction
  \State validate Eq.~\eqref{eq:editbound} and rubric items 1--5 before querying the answer head
  \If{$R^{(j)}$ is admissible}
    \State add $j$ to $J_{\mathrm{val}}$; then measure answer response $\Delta_j$
  \EndIf
\EndFor
\If{$|J_{\mathrm{val}}|<m_{\min}$} \State \Return \textsc{Abstain} with intervention-coverage flag \EndIf
\State compute $S_{RA}$ using Eq.~\eqref{eq:support}
\State sample $\{(R_k,A_k)\}_{k=1}^{K}$ and normalize answers
\State compute $C_A$ and $D_R$ using Eqs.~\eqref{eq:consensus}--\eqref{eq:stability}
\State $s\gets100\exp(\sum_m w_m\log(s_m+\epsilon))$
\State state $\gets$ calibrated \textsc{Accept}, \textsc{Review}, or \textsc{Abstain}
\State \Return $(s,\text{state},t^*)$
\end{algorithmic}
\end{algorithm}

\section{Detection Architecture}
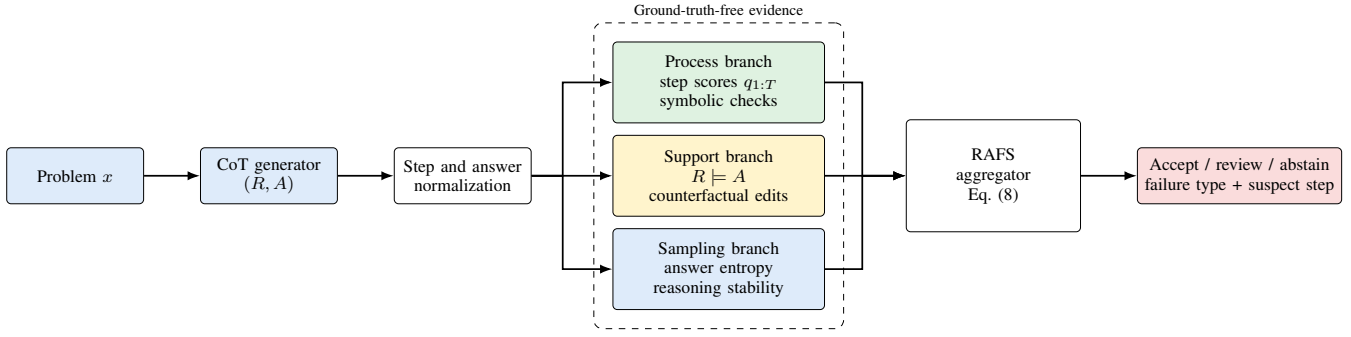
\begin{figure*}[t]
\centering
\resizebox{0.98\textwidth}{!}{%
\begin{tikzpicture}[
  font=\footnotesize,
  box/.style={draw,rounded corners=2pt,align=center,minimum height=9mm,minimum width=22mm,fill=white},
  branch/.style={box,minimum width=34mm,minimum height=13mm},
  arr/.style={-{Latex[length=2mm]},thick},
  node distance=7mm and 9mm]
  \node[box,fill=coolblue] (input) {Problem $x$};
  \node[box,right=of input,fill=coolblue] (gen) {CoT generator\\$(R,A)$};
  \node[box,right=of gen] (parse) {Step and answer\\normalization};
  \node[branch,right=13mm of parse,yshift=15mm,fill=goodgreen] (proc) {Process branch\\step scores $q_{1:T}$\\symbolic checks};
  \node[branch,right=13mm of parse,fill=warnyellow] (support) {Support branch\\$R\models A$\\counterfactual edits};
  \node[branch,right=13mm of parse,yshift=-15mm,fill=coolblue] (sample) {Sampling branch\\answer entropy\\reasoning stability};
  \node[box,right=13mm of support,minimum width=28mm,minimum height=18mm,fill=white] (agg) {RAFS\\aggregator\\Eq.~\eqref{eq:rafs}};
  \node[box,right=of agg,minimum width=31mm,fill=badred] (decision) {Accept / review / abstain\\failure type + suspect step};
  \draw[arr] (input)--(gen);
  \draw[arr] (gen)--(parse);
  \draw[arr] (parse.east)--++(5mm,0)|-(proc.west);
  \draw[arr] (parse)--(support);
  \draw[arr] (parse.east)--++(5mm,0)|-(sample.west);
  \draw[arr] (proc.east)--++(6mm,0)|-(agg.west);
  \draw[arr] (support)--(agg);
  \draw[arr] (sample.east)--++(6mm,0)|-(agg.west);
  \draw[arr] (agg)--(decision);
  \node[draw,dashed,rounded corners,fit=(proc)(support)(sample),inner sep=3mm,label=above:{\scriptsize Ground-truth-free evidence}] {};
\end{tikzpicture}
}
\caption{RAFS detection architecture. Independent evidence branches make the final decision auditable and allow deterministic tools or learned verifiers to be exchanged without changing the metric.}
\label{fig:pipeline}
\end{figure*}

Figure~\ref{fig:pipeline} shows the operational pipeline, complementing direct reasoning-path evaluation with explicitly separated evidence branches \cite{bao2024direct}. The generator first produces a primary trace and answer. A parser separates atomic claims, calculations, and the answer span. Three analysis branches then run in parallel: (i) a process verifier scores each prefix and invokes symbolic tools when possible; (ii) an answer-support module tests entailment and counterfactual dependence; and (iii) a sampler estimates answer entropy and within-answer reasoning stability. The aggregator returns RAFS together with decomposed scores, so a user can tell whether rejection arose from a local invalid step, a trace--answer mismatch, or broad sampling uncertainty.

The pipeline has two safeguards. First, \emph{cross-family independence} means that no judge, edit validator, or answer head shares a base-model family or instruction-tuning lineage with the trace generator: for example, DeepSeek judges Llama traces, Llama judges Mistral traces, and Mistral judges DeepSeek traces. Merely changing a system prompt, adapter, quantization, or checkpoint size within Llama does not count as independent. Symbolic checks are family-independent. The confirmatory ablation holds prompts, traces, calibration size, and decision thresholds fixed and compares (i) same-checkpoint, (ii) different checkpoint in the same family, and (iii) cross-family judging; it reports silent-failure AUPRC, recall at 90\% precision, calibration, and judge--generator error correlation. Second, all critic probabilities are calibrated on trace- and step-level labels. If symbolic execution is available, it overrides a conflicting soft arithmetic judgment; for open-ended claims, disagreement increases uncertainty rather than being resolved by unjustified majority vote.

\section{Experimental Design}
The confirmatory evaluation uses GSM8K and MATH as complementary arithmetic
and competition-mathematics benchmarks \cite{cobbe2021gsm8k,hendrycks2021math}.
\subsection{Feasibility Pilot and Confirmatory Boundary}
Before freezing the confirmatory study, we reserve 100 disjoint items (50 GSM8K and 50 MATH) and one 7--8B generator for an implementation-only pilot. The pilot must archive every emitted trace, atomic segmentation, calibrated $q_t$, intervention candidate and rejection code, answer-head response, component score, latency, and human quadrant label. Its mandatory report includes the median and interquartile range of $q_t$ (plus the lower-tail mass below $\tau_v$), candidate admission by each rubric item, the instance fraction with $|J_{\mathrm{val}}|<2$, and the four-quadrant counts with exact binomial intervals. It also compares geometric, arithmetic, minimum-gated, and logistic aggregation with nested cross-validation internal to the pilot. Since these comparisons are descriptive, they don’t alter Eq.~\eqref{eq:rafs} or the confirmatory hypotheses.

The current source manuscript does not contain pilot artifacts or executable model outputs that allow verification of such quantities. Consequently, we do not produce numerical 'pilot results' based on illustrative scores. This paragraph will be replaced with archived data and the citation of the item-level evidence table if this version is to be claimed as having the feasibility evidence. Until that time, the pilot serves as a pre-confirmatory gate rather than as evidence. It is better to have this clear limitation, rather than allow the hand-assigned values of Fig.~\ref{fig:example} to be mistaken for observations.

\subsection{Models and Benchmarks}
We suggest assessment of three families of open-weight models with different training recipes and capabilities: Llama 3.1/3.2 Instruct, Mistral 7B or Mixtral Instruct, and DeepSeekMath-7B or a distilled DeepSeek-R1 variant \cite{deepseekmath2024, deepseekr1}. Models are assessed in their publicly available inference configurations at 7–8B with the option of including a larger checkpoint to assess scaling. The primary data sets are GSM8K and MATH, which feature problem solving for multi-step grade school arithmetic and advanced topics including algebra, counting, geometry, intermediate algebra, number theory, precalculus, and probability, presented at a competition level of difficulty.

For each problem, the primary (R, A) evidence is generated via greedy decoding, and the evidence is generated via K = 16 samples with temperature sampling. Answers are normalized via task-specific parsers. Reasoning is segmented at the boundaries of sentences and transitions of equations, and is then audited manually for a stratified subset. Experiments use fixed prompts, decoding budgets, and random seeds across models. No benchmark solution or gold answer is exposed to RAFS at inference; references are used only for post-hoc accuracy and quadrant labels.

\subsection{Four-Quadrant Annotation and Power}
We increase the annotation target from 1,200 to 2,400 responses, stratified across model, dataset, and answer correctness. Two mathematically qualified annotators independently identify the earliest invalid step, legitimate recovery, trace--answer entailment, and quadrant; a third adjudicates disagreements. We report Cohen's $\kappa$ for reasoning validity and Krippendorff's $\alpha$ for error location. To reduce hindsight bias, annotators first inspect $R$ with the answer span masked, then judge $R\models A$ after unmasking it.

The increase is driven by the expected rarity of silent failures. At 10\% prevalence, $n=1{,}200$ yields only about 120 positives and unstable tail precision; $n=2{,}400$ yields about 240. A preregistration-time Monte Carlo calculation (10\% prevalence, baseline AUPRC $0.30$, paired score correlation $0.60$, one-sided paired bootstrap test at $\alpha=.05$) gave 81\% power for an AUPRC increase of $0.068$ over 500 simulated studies (200 bootstrap replicates, seed 19). These distributional assumptions are planning inputs, not pilot findings. If the feasibility pilot's exact 95\% prevalence interval implies fewer than 200 silent positives in 2,400 annotations, sampling is adaptively enriched with answer-correct/low-process-score cases until at least 200 silent positives are adjudicated; inverse-probability weights recover population AUPRC. The enrichment rule and weights are fixed before confirmatory labels are opened.

\subsection{Baselines and Ablations}
RAFS is compared with: (1) exact-match answer accuracy; (2) answer self-consistency; (3) mean LLM-as-a-judge step score; (4) ROSCOE-style reference-free metrics; (5) a PRM score; and (6) entailment between the full chain and answer. Ablations remove $P_R$, $I_{RA}$, $C_A$, or $D_R$; replace the geometric mean with arithmetic, hard-minimum, and learned logistic aggregation; vary $K\in\{4,8,16,32\}$; and compare same-checkpoint, within-family, and cross-family judges. A targeted ablation compares random perturbations with pivotal-step interventions to test whether semantic precision is necessary.

\subsection{Evaluation Metrics and Hypotheses}
The primary endpoint is macro-F1 over the four quadrants. Because silent failures are the central risk, we also report their area under the precision--recall curve (AUPRC), recall at 90\% precision, and expected calibration error. Step localization is measured by exact earliest-error accuracy and distance from the expert index. Selective prediction is evaluated by risk--coverage curves: as low-RAFS cases are deferred, the error rate among accepted responses should decrease.

We test four preregistered hypotheses. \textbf{H1}: answer accuracy overestimates valid reasoning by merging silent failures with faithful successes. \textbf{H2}: RAFS improves silent-failure AUPRC over answer consensus and mean step score. \textbf{H3}: counterfactual sensitivity contributes most strongly when answers are correct but reasoning is invalid. \textbf{H4}: conditional reasoning stability detects shared-answer rationalization that answer entropy alone misses. Confidence intervals are obtained by problem-level bootstrap; paired model comparisons use permutation tests with Holm correction.

\begin{table}[t]
\caption{Planned reporting matrix. No numerical result is asserted before execution of the preregistered experiments.}
\label{tab:reporting}
\centering
\footnotesize
\begin{tabular}{@{}lccc@{}}
\toprule
Method & 4-way F1 & Silent AUPRC & ECE $\downarrow$ \\
\midrule
Exact match & -- & baseline & -- \\
Answer self-consistency & planned & planned & planned \\
Step judge / PRM & planned & planned & planned \\
Entailment only & planned & planned & planned \\
RAFS (ours) & planned & planned & planned \\
\bottomrule
\end{tabular}
\end{table}

\section{Coverage, Cost, and Deployment Modes}
\subsection{Intervention Coverage and Expected Abstention}
The five-item rubric intentionally sacrifices coverage for easily interpretable interventions. Before pilot measurements exist, deployment capacity will rely on sensitivity analysis, not reported empirical rates. If we attempt to admit three pivotal candidates with each candidate assuming a 55 percent admission probability, the candidate rejection rate is 45 percent. The binomial probability of two or fewer admissions is then calculated to be 42.5 percent. Attempting to admit five independent candidates lowers this planning parameter to 13.1 percent. Independence is a loose assumption; the correlated validator failures may result in a higher abstention rate. Therefore, this feasibility pilot provides both candidate and instance coverage, with bootstrapped rejection codes for locality, edit distance, semantic isolation, leakage, parser failure, validator disagreement, and ambiguous direction, with proposed abstention rates determined to be a 40 percent rejection rate for selective audits of a small number of high-consequence mathematical solutions. This abstention rate is not acceptable for an always-on benchmark scorer. Operational policy will attempt to admit candidates sequentially, up to five. If two admissions are realized, the policy will apply a coverage abstention, as will the policy if the coverage threshold is not met. Coverage abstention will be distinguished from a low-score review, in that coverage abstention indicates a lack of evidence for causation, not unfaithfulness of the trace.

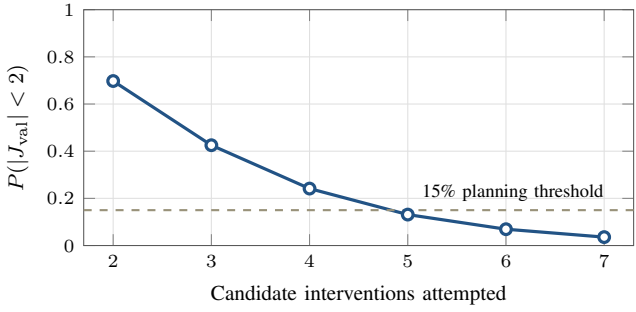
\begin{figure}[t]
\centering
\begin{tikzpicture}
\begin{axis}[
  width=\columnwidth,
  height=4.7cm,
  xmin=1.7,xmax=7.3,
  ymin=0,ymax=1,
  xtick={2,3,4,5,6,7},
  ytick={0,0.2,0.4,0.6,0.8,1},
  yticklabel style={/pgf/number format/fixed,
    /pgf/number format/precision=1},
  xlabel={Candidate interventions attempted},
  ylabel={$P(|J_{\mathrm{val}}|<2)$},
  tick label style={font=\scriptsize},
  label style={font=\footnotesize},
  grid=major,
  major grid style={draw=black!12},
  axis line style={black!65}]
  \addplot[darkblue,very thick,mark=*,mark size=2pt,
    mark options={fill=white}] coordinates {
      (2,0.6975) (3,0.4253) (4,0.2415)
      (5,0.1312) (6,0.0692) (7,0.0357)};
  \addplot[warnyellow!60!black,thick,dashed,domain=1.7:7.3] {0.15};
  \node[font=\scriptsize,anchor=south west] at (axis cs:5.05,0.15)
    {15\% planning threshold};
\end{axis}
\end{tikzpicture}
\caption{Planning sensitivity of coverage abstention to the candidate budget,
assuming independent admission probability $0.55$ and $m_{\min}=2$.
This is an analytical planning curve, not an observed pilot result.}
\label{fig:coverage}
\end{figure}

\subsection{Call and Latency Budget}
Consider the defaulted $K=16$. Three candidate edits, two validators per edit, and two admitted interventions. As an example, let us consider a single primary solution. Computing k=16 resamples will likely involve a batched step-verification call, support/pivotal-analysis calls, edit generations, validations, and counterfactual answer-head calls. Note that prefix-by-prefix verification (without batching) requires T - 1 calls, where T is the total number of tokens in your answer. Most of the calls within the sampling and validation branches can be executed in parallel. Therefore, the total (wall-clock) time of the latency will be determined by four to seven waves of calls, rather than 31. During the sampling and validation calls within RAFS, token consumption (estimation) is expected to be between 12 to 25 times greater than a base inference, as many judge outputs are relatively short. The pilot measures the uncached tokens, along with median/95th latency, along with the system's hardware, batch size, and peak memory measures.

Full RAFS is designed for expert review and dataset auditing. For these applications, the cost of a missed silent failure is greater than dozens of short verifier calls. \emph{RAFS-lite} uses $K=4$, one batched process judgment and trace--answer entailment, and the same NLI stability calculation, but skips counterfactual generation and reports $(P_R,e,C_A,D_R)$ with a separately calibrated lite score. This version retains no claims of counterfactual faithfulness and consists of seven calls, at a minimum. Ambiguous and high-impact cases are to be escalated to full RAFS. A basic screening mode which performs only symbolic checks and answer normalization is to be used in lieu of full RAFS.

\section{Illustrative Silent-Failure Case}
\begin{figure}[t]
\centering
\begin{tikzpicture}[font=\scriptsize,
  line/.style={draw,rounded corners,align=left,text width=6.7cm,inner sep=2.2mm},
  arr/.style={-{Latex[length=1.6mm]},thick}]
  \node[line,fill=coolblue] (q) {\textbf{Question:} An item is bought for \$40 and sold for \$55. What is the profit?};
  \node[line,below=2mm of q] (s1) {Step 1: Profit equals selling price minus purchase price. \hfill $q_1=.98$};
  \node[line,below=2mm of s1,fill=badred] (s2) {Step 2: $55-40=10$. \hfill $q_2=.03$};
  \node[line,below=2mm of s2,fill=badred] (s3) {Step 3: Add a \$5 ``transaction adjustment,'' giving \$15. \hfill $q_3=.08$};
  \node[line,below=2mm of s3,fill=goodgreen] (a) {\textbf{Final answer:} \$15. \hfill Exact match: \checkmark};
  \node[draw,thick,rounded corners,below=2mm of a,align=center,fill=warnyellow,text width=6.7cm,inner sep=2mm] (out) {$P_R$ low; $R\models A$ unsupported; intervention unstable\\\textbf{RAFS: Review---silent-failure candidate at Step 2}};
  \draw[arr] (q)--(s1); \draw[arr] (s1)--(s2); \draw[arr] (s2)--(s3); \draw[arr] (s3)--(a); \draw[arr] (a)--(out);
\end{tikzpicture}
\caption{Illustrative silent failure. The numerical RAFS is intentionally omitted until weights and verifier outputs are calibrated; the decomposed evidence already identifies the failure mode.}
\label{fig:example}
\end{figure}
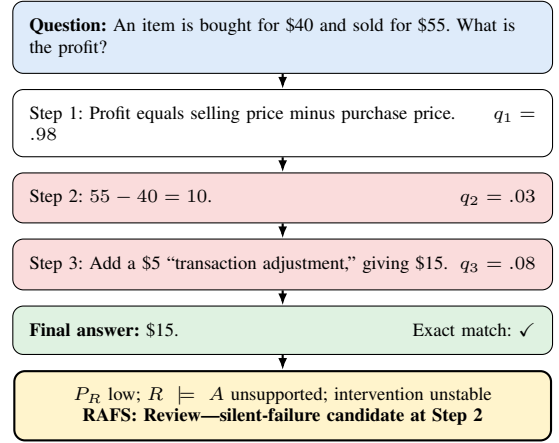

Figure~\ref{fig:example} illustrates a failure that exact match cannot expose. The problem asks for profit after buying an item for \$40 and selling it for \$55. The trace incorrectly computes $55-40=10$, then introduces an unsupported ``\$5 adjustment'' and outputs \$15. The answer happens to match the reference, but neither local validity nor global entailment is acceptable. A pivotal-step intervention that corrects the subtraction eliminates the need for the invented adjustment; the model's original answer mechanism is therefore unstable. RAFS flags the trace even though $Y_A=1$.

This example also shows why an answer-only perturbation is insufficient. If a judge sees the correct reference, it may overlook the invalid chain through outcome bias. RAFS instead requires agreement among local checks, trace-level support, and intervention behavior. In a real evaluation, the displayed $q_t$ values would be calibrated verifier outputs rather than hand-assigned illustrative values.

\section{RAFS-Guided Correction}
Detection is helpful only when the uncertainty can be acknowledged and corrected. We introduce the correction operator $\mathcal{C}(x,R,A,\boldsymbol{s})$, where $\boldsymbol{s} = (P_R,S_{RA},C_A,D_R)$, which acts the least selectively based on the supporting diagnostic evidence. The operator subliminally provides the final answer to a new solver only when answer comparison is made, making anchoring to the potentially incorrect response less severe.

\subsection{Localized Reasoning Repair}
When $P_R$ is low and an earliest suspect step $t^*$ is identified, the
system preserves the verified prefix $r_{1:t^*-1}$ and regenerates the suffix
under the requirement that replacement steps state their premises and satisfy
available symbolic checks. We draw $B$ candidate suffixes and reject any
candidate whose first replacement step fails verification. The highest-RAFS
surviving pair is selected only if it improves on the original by a margin
$\delta$ and reaches the accept threshold; otherwise, the system abstains.
Formally,
\begin{equation}
(R',A')=\arg\max_{(\tilde R,\tilde A)\in\mathcal{B}(t^*)}
\mathrm{RAFS}(x,\tilde R,\tilde A),
\end{equation}
subject to $r'_{1:t^*-1}=r_{1:t^*-1}$ and $q'_{t^*:T}\geq\tau_v$. Preserving the verified prefix limits unnecessary semantic drift and makes the correction auditable.

\subsection{Answer-Extraction Repair}
When process and entailment scores are high but sample answers differ, we consider failures post-derivation. The system freezes R, and a constrained answer extractor is invoked. The output is required to be a deterministic function of the final supported claim. For multiple-choice tasks, this directly includes a value-to-option mapping check. For numerical tasks, checks for sign, units, simplifications, and rounding are included. Most importantly, we do not regenerate reasoning, as this has the potential to replace a valid derivation with a less faithful one. If the corrected A is entailed by R and increases answer consensus, we log this as an extraction correction and not a reasoning improvement.

\subsection{Independent Re-Solving and Abstention}
A low SRA with high answer consensus is a typical example of a “silent-failure” behavior. Local edits can be risky because the answer may end up being a binding constraint to all candidate repairs. The system, therefore, employs a different prompt, and checkpoint, and/or a different symbolic route with both R and A concealed, and requests an independent solution.Agreement is determined for the intermediate claims and is considered before answer comparison. If the independent solution is achieved via the same validated route and answer,it replaces the initial explanation. If it is via a different validated route and answer, the case is escalated.
\subsection{Correction Evaluation}
Correction is evaluated separately from detection. Metrics include the correction success rate, which is the rate of previously invalid cases which transitioned to faithful successes; the regression rate, which is the rate of faithful successes which were negatively impacted; edit locality, which is measured in the normalized length of the suffix which was edited; and compute overhead. We also consider false reassurances, which are cases in which RAFS increased, but the validity of the reasoning from the expert did not improve.A valid correction must improve the reasoned expert quadrant,not merely optimize the automated score. Other metrics included full-response regeneration, self-refine prompting,answer-only re-extraction, and the proposed failure-type-specific policy. This design is meant to test if RAFS presents an actionable structure, beyond merely ranking the responses.

\section{Analysis and Discussion}
\subsection{Interpreting RAFS}
RAFS is best seen as a formalized structured warning signal. A high score denotes that the observed trace is locally credible, supports the answer, and is stable across the defined sampling and intervention distribution. While a high score verifies the observed trace is stable, it does not support the veracity of the answer, as model family and model verifier can have a shared coherent fallacy. Low scores are more prescriptive. The described components determine whether a user should repair that step, re-run the answer extraction, engage an external solver, or abstain from the task.

The four-quadrant model introduces a new dimension to model comparison. Consider two systems having the same exact-match accuracy. A human underlining system that produces fewer incorrect silent failures is, in general, superior, given that its correct responses are more frequently backed up. A system that exhibits a higher number of reasoning false negatives may be process competent and may have a sufficiently capable answer head or a sufficiently constrained answer decoder.

RAFS makes these distinctions clear, given that accuracy has a scoring system that has an equal distribution across the systems.

\subsection{Contribution to Explainable and Safe AI}
Measurable limitations on explainable AI show where explainable AI can be advanced. RAFS evaluates dependency between decision and explanation and develops human oversight. A reviewer could see why a decision is suspicious and if the decision is insensitive to rationale changes. Selective deployment can apply more powerful models, expert systems, and symbolic tools to unreasonably low scoring cases while keeping automation for stable and predictable cases.

RAFS can produce multi-objective feedback. During training, silent failures should not be rewarded in the same way as faithful successes just because the two answers are the same. Reasoning false negatives that do not have an obvious outcome should also not be rewarded in the same way as evident failures. In the future, a trainer could reach an answer that is correct in the RAFS sense, optimizing score correctness while achieving suboptimal RAFS, preventing outcome from dominating process.. To avoid reward hacking, however, RAFS should be periodically audited with novel intervention templates and independent human labels.

\subsection{Adversarial Inference-Time Gaming}
Inference-time gaming is distinct from training-time reward hacking. Recent analyses of reasoning-model disclosure and naturally occurring CoT faithfulness motivate treating this as a separate evaluation threat \cite{chen2025reasoning,xu2025wild}. A generator that knows the audit may emit verbose, redundant rationalizations whose claims are difficult to atomize, or write an answer-independent explanation engineered to remain superficially coherent under familiar edits. Such behavior can raise judge confidence while reducing intervention admissibility, producing abstention rather than a clean low score. It can also repeat the same claim in several forms so that a local edit appears not to affect the answer.

Mitigations include hiding the sampled intervention family from the generator; drawing periodic novel templates that change numeric, relational, and proof-obligation claims; collapsing semantically duplicate claims before selecting pivotal steps; and red-teaming with models fine-tuned to maximize RAFS while preserving invalid solutions. Results are reported both on standard traces and on this adaptive adversarial set. A rise in intervention rejection is itself a monitored signal, and coverage abstention cannot be converted to acceptance by verbosity. These measures reduce predictable gaming but do not establish strategy-proofness.

\subsection{Limitations and Threats to Validity}
First, natural-language steps are difficult to atomize; a single sentence can contain several claims, and an apparent error can be repaired later. Second, verifier bias can correlate with generator bias. Independent ensembles and symbolic checks mitigate but do not eliminate this problem. Third, counterfactual edits may create unnatural chains, so intervention validity must be screened. Fourth, reasoning diversity is not always bad: two distinct proofs can support the same answer. Our conditional stability representation must compare normalized claims or proof obligations rather than surface wording.

Fifth, RAFS evaluates emitted text, not private activations. Even strong counterfactual response is only evidence that the text mediates the tested prediction pathway. Sixth, the planning calculation for abstention assumes independent candidate admissions, whereas rubric failures may be correlated. Finally, all present definitions, validators, and claims are restricted to GSM8K/MATH-style mathematical reasoning.

\section{Conclusion}
Final-answer accuracy alone cannot determine whether a mathematical CoT is valid. It hides correct answers supported by invalid chains and misclassifies valid derivations followed by answer-selection errors. We formalized this reasoning--answer consistency gap and introduced RAFS, a reference-free score combining step validity, trace-to-answer support, counterfactual sensitivity, answer consensus, and conditional reasoning stability. The pipeline produces an auditable failure type and abstains under uncertainty or inadequate intervention coverage. The preregistered Llama, Mistral, and DeepSeek study on GSM8K and MATH will test whether RAFS improves silent-failure detection over accuracy, self-consistency, and isolated process scores; until the feasibility artifacts and confirmatory results exist, the contribution is a technically specified and falsifiable mathematical-reasoning framework, not a demonstrated safety guarantee.

The component decomposition interprets operational failures and lends itself to more than just a single score. Analyzing process validity can encourage step repair. Low support predictability can encourage independent re-solving. Low support sampling can encourage review or abstention. Most importantly, RAFS captures unsupported correct answers as a unique evaluation outcome. In the absence of RAFS, unsupported correct answers are captured as part of overall correct answer submissions.

This framework allows for clear analysis of various interventions and the uncertainty they produce. Because unsupported evidence cannot be confused with positive evidence, this framework promotes evidence value evaluation. This framework allows for clear analysis for interventions with a low final answer score, but a great deal of rational support.

\section{Future Work}
Next, we will execute and publish the feasibility pilot, followed by the preregistered confirmatory experiments spanning model scales, decoding strategies, and mathematical difficulty. Any extension to scientific reasoning, medicine, or law will likely need new domain-specific verifiers, retrieval or evidence grounding, expert validity definitions, and new calibration, all of which are absent here. Future work could address the atomization, adaptive interventions, and ensembles of verifiers, and apply adversarially optimized generators. Longitudinal and human review studies could evaluate if RAFS-guided training and calibrated abstention help improve the quality of mathematical decisions.

It will be crucial to examine transfer to new problem templates, new proof styles, and significantly longer proofs. Time-efficient implementations should be able to achieve almost full RAFS reliability at the cost of RAFS-lite using adaptive sampling, verifier cascades, and early stopping. Constructing public intervention toolkits, and standardized reporting frameworks will help to evaluate, in a consistent manner, adversarial robustness, verifier families, and dimensions of coverage, calibration, and latency.

New benchmarks should also involve controlled silent-failure with goal-directed reasoning and answer extraction tasks with designed failures and self-correcting errors. Such challenge sets would enable repeatable stress testing and reveal whether improvements persist outside naturally occurring benchmark errors.

Any later extension to autonomous cloud services or critical-infrastructure
cyber defence would require domain-specific state, evidence, and intervention
validators beyond the present mathematical setting
\cite{shukla2026agentic,atul2025federated}.

\end{document}